\documentclass[prl,floatix,amssymb,showpacs,amsmath,superscriptaddress]{revtex4-1}
\usepackage{graphicx}
\usepackage{epsfig,color}
\usepackage{amsmath,amssymb}
\begin{document}
\title{Mapping curved spacetimes into Dirac spinors}

\date{\today}

\author{Carlos Sab{\'{i}}n}
\affiliation{Instituto de F{\'i}sica Fundamental, CSIC,
Serrano, 113-bis,
28006 Madrid (Spain)}
\email{csl@iff.csic.es}

\begin{abstract}
We show how to transform a Dirac equation in a curved static spacetime into  a Dirac equation in flat spacetime. In particular, we show that any solution of the free massless Dirac equation in a 1+1 dimensional flat spacetime can be transformed via a local phase transformation into a solution of the corresponding Dirac equation in a curved static background, where the spacetime metric is encoded into the phase. In this way, the existing quantum simulators of the Dirac equation can naturally incorporate curved static spacetimes. As a first example we use our technique to obtain solutions of the Dirac equation in a particular family of interesting spacetimes in 1+1 dimensions.
\end{abstract}
\maketitle

The growing interest in quantum simulators is not restricted to their potential role as non-universal post-classical computers.
They are also powerful tools to explore open problems in the frontiers of theoretical physics and even to pose questions beyond them.

Along these lines, quantum simulators of relativistic quantum mechanical dynamics have received much attention in the last years. Originally intended as a fundamental description of fermionic elementary particles, relativistic quantum mechanics is considered nowadays as a single-particle approximation to Quantum Field Theory. Thus it is an accurate description of fermionic particles in the low-energy regime below the pair-creation energy threshold.  Thanks to cutting-edge quantum simulators, several hitherto unobserved aspects of the Dirac (\textit{Zitterbewegung}, Klein paradox etc.\cite{zitt1,klein1}) and Majorana (antiunitary transformations, exotic dynamics etc. \cite{majorana}) equations in 1+1 dimensions have been implemented in the laboratory in the most advanced quantum platforms, such as trapped ions \cite{zitt2,klein2,experiment1}, optical waveguide arrays \cite{majorana2} or cold atoms \cite{salger2011klein}, while theoretical proposals with superconducting circuit architectures are also available \cite{julen}. 

Recently, a theoretical proposal for a quantum simulator of the Dirac equation in curved 1+1 dimensional spacetimes by means of  coupled waveguide arrays has been introduced \cite{diraccurved1}, extending the interest in relativistic quantum mechanical simulators beyond the realm of flat-spacetime physics. This equation can be seen as a single-particle approximation to quantum field theory in curved spacetime, which in turn is itself regarded as a sound intermediate step to the final theory of quantum gravity, operating in a regime where backreaction effects from the spacetime are neglected.

In this work, we show how to transform a Dirac equation in curved spacetime into a Dirac equation in flat spacetime. In particular, we show that any solution of a free massless Dirac equation in 1+1 dimensions can be transformed into a solution of the same equation in curved spacetime, for a broad generic class of spacetime metrics. This transformation is just a local phase, which carries the information of the particular spacetime metric being considered. Thus, all the aforementioned multiplatform quantum simulators - which are able to reproduce a free massless 1+1 Dirac equation- are immediately able to analyse curved spacetime metrics as well, without the need of actually simulating a new Hamiltonian. Instead, the new physics can be obtained by applying our transformation technique to the solutions of the corresponding flat-spacetime Hamiltonian.

As an example of interest, we show how to apply our technique to a family of 1+1 dimensional spacetimes containing traversable wormholes.  
Although it seems that these objects do not appear naturally in our Universe \cite{search},  wormholes lie at the edge of theoretical physics and defy our understanding of key physical principles, such as causality. For these reasons, they have always attracted much attention from a foundational  and pedagogical viewpoint  \cite{morristhorne, morristhorne2} and they are the subject of current intense research \cite{taylor, geodesics,shadows}, including possible simulations \cite{magneticwormhole, rousseax,qswh}. We use our technique to obtain for the first time analytical solutions of the Dirac equation in the presence of a traversable wormhole, by applying a local phase transformation to a wavepacket solution of the corresponding flat spacetime equation. In this way, we are able to see how the probability density of a Dirac particle is distorted by the presence of a wormhole, and specifically when traversing its throat. The same techniques can be applied to numerical solutions or to the experimental data collected in real quantum simulators, adding an extra element to the toolbox of quantum simulators of relativistic quantum mechanics. Extensions to other 1+1 spacetimes of interest would be straightforward. 

\section*{Methods}
  
We  will start the description of our results by considering  a free Dirac equation in the background of a curved spacetime in 1+1 dimensions ($\hbar=c=1$) \cite{diraccurved1}:
\begin{equation}
i(\partial_t+\frac{\dot{\Omega}}{2\Omega})\psi=(-i\,\sigma_x (\partial_x+\frac{\Omega'}{2\Omega})+\sigma_z\,\Omega m)\psi, \label{eq:diraccurve}
\end{equation}
where $\dot{}$ and $'$ denote time and space partial derivatives respectively, and $\sigma_{x,z}$ are Pauli matrices:  
$\sigma_x=\begin{pmatrix}0 & 1 \\ 1 & 0\end{pmatrix},\,\sigma_z=\begin{pmatrix}1 &0 \\ 0 & -1\end{pmatrix}.$
Here $\Omega$ carries all the information of the spacetime background, since in 1+1 dimensions the spacetime metric can always be brought to the generic form:
\begin{equation}
ds^2=\Omega^2(x,t)(dt^2-dx^2).\label{eq:metric}
\end{equation}
The details on the derivation can be found for instance in \cite{diraccurved1}.
In the case of a massless field $m=0$ and a static spacetime $\dot{\Omega}=0$, Eq. (\ref{eq:diraccurve}) becomes:
\begin{equation}\label{eq:diraccurve2}
i\partial_t\psi=(-i\,\sigma_x \partial_x+V(x)\sigma_x)\psi,
\end{equation}
where all the effect of the background metric is captured by the effective non-hermitian potential:
\begin{equation}\label{eq:potential}
V(x)=-i\frac{\Omega'}{2\Omega}.
\end{equation}

\section*{Results}

Now, we will exploit a technique \cite{potwopot}, which transforms Dirac equations including potentials into free equations. Let us consider the local phase transformation:
\begin{equation}\label{eq:transformation}
\psi=e^{-i\int{V(x)\,dx}}\phi= \Omega^{-1/2}\phi,
\end{equation}
where we have used Eq. (\ref{eq:potential}).
Introducing Eq. (\ref{eq:transformation}) into Eq. (\ref{eq:diraccurve2}), we see that an extra term coming from the spatial derivative cancels out the effective potential, yielding the following equation for $\phi$:
\begin{equation}\label{eq:diraccurve3}
i\partial_t\phi=-i\,\sigma_x \partial_x\phi,
\end{equation}
which is a free massless Dirac equation in a 1+1 dimensional flat spacetime. Therefore, any solution of the free flat-spacetime Dirac equation can be converted into a solution of the Dirac equation in curved spacetime by means of a local phase transformation. Of course, this does not amount to say that both dynamics are equivalent- otherwise, the effect of the spacetime would be trivial. Indeed, since $V(x)$ is an imaginary potential, the local phase is real and the probability densities 
for $\psi$ and $\phi$ are different. In particular, by combining Eqs. (\ref{eq:potential}) and (\ref{eq:transformation}) we find:
\begin{equation}\label{eq:densities}
|\psi|^2= \Omega^{-1}|\phi|^2.
\end{equation}

Thus, we have shown that in order to simulate the non-trivial effects of a curved spacetime into the dynamics of Dirac particle, it is not necessary to engineer an extra term in the Hamiltonian -in particular, a non-hermitian potential depending on the metric- but only to consider a particular phase transformation of a free Dirac spinor. In this way, any quantum simulator of a flat-spacetime Dirac equation is immediately transformed into a quantum simulator of a curved-spacetime Dirac equation.

Let us now focus on a particular example of interest. We will analyse a family of spacetimes containing a traversable wormhole.

A traversable 1+1 dimensional massless wormhole spacetime can be characterised by \cite{morristhorne2,qswh}
\begin{equation}
ds^2=-dt^2+\frac{1}{1-\frac{b(r)}{r}}\,dr^2, \label{eq:metric}
\end{equation}
where the shape function $b(r)$ is a function of the radius $r$ only. There is a singular point of $r$ at which $b\, (r=b_0)=r=b_0$, which determines the position of the wormhole's throat and defines two different Universes or two asymptotically flat regions within the same Universe -as $r$ goes from $\infty$ to $b_0$ and then back from $b_0$ to $\infty$.  
The properties of the wormhole will depend on the form of the shape function $b(r)$. 

Of particular interest is the following family of wormholes \cite{ellis,morristhorne, geodesics,taylor,qswh}: 
\begin{equation}\label{eq:example}
b(r)=\frac{b_0^2}{r}.
\end{equation}
It is shown in \cite{qswh} that this family of spacetimes  is totally equivalent to the  following one:
\begin{equation}\label{eq:example2}
ds^2=-(1-\frac{b(r)}{r})\,dt^2+\,dr^2,
\end{equation}
which are spacetimes in which the speed of light depends on the radius $r$ according to:
$c^2(r)=(1-\frac{b(r)}{r})$. We can also introduce a new coordinate
\begin{equation}\label{eq:changecoord}
x=\int{\frac{dr}{c(r)}}=\pm\sqrt{r^2-b_0^2},
\end{equation}
where $x>0$ and $x<0$ correspond to the two different regions at both sides of the throat. This brings finally the spacetime metric to the form:
\begin{equation}\label{eq:example}
ds^2= c^2(x)(-dt^2+dx^2).
\end{equation}

We can thus exploit our technique by noticing -using Eq. (\ref{eq:metric})- that:
\begin{equation}\label{eq:Omegaexample}
\Omega^2(x)=c^2(x)=1-\frac{b_0^2}{x^2+b_0^2}.
\end{equation}
Then, we find that:
\begin{equation}\label{eq:omegaprimeoveromega}
\frac{\Omega'}{\Omega}=\frac{b_0^2}{x(b_0^2+x^2)}.
\end{equation}
\begin{figure*}[t!] 
\hspace*{-0.1cm}
\includegraphics[width=\textwidth]{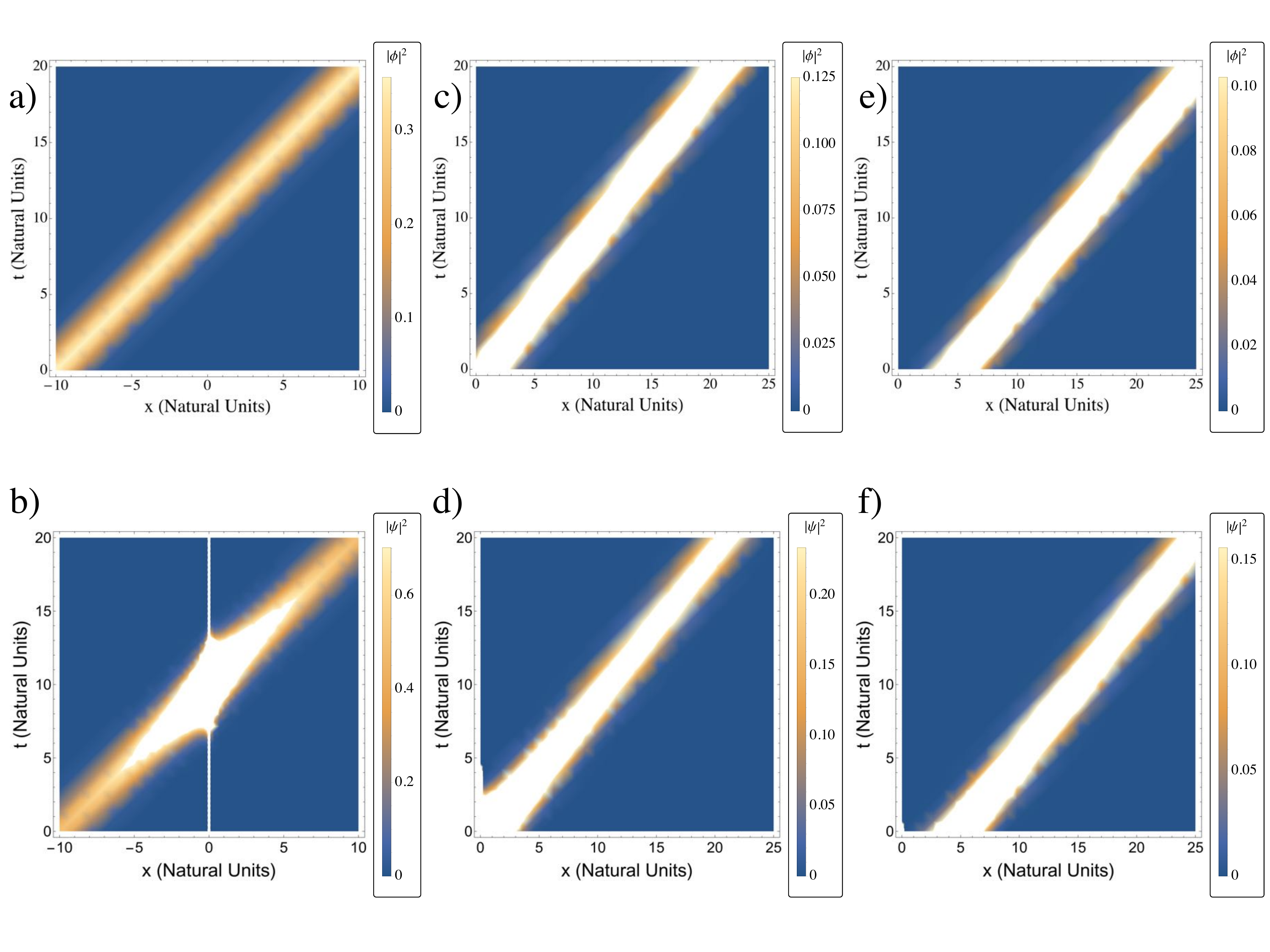}
\caption{Probability density dynamics of a Dirac wavepacket in: a), c), e) (top row) flat spacetime and b), d), f) (bottom row) wormhole spacetime with $b_0=10$. The initial position is: a), b) (first column) $x_0=-10$; c), d) (second column) $x_0=1$ and e), f) (third column) $x_0=5$. The Gaussian width is $\sigma=5$ in all cases. Note that the point $x=0$ is not included in b).}
\label{fig:plots}
\end{figure*}
Thus the Dirac equation in 1+1 dimensions of a massless particle in a spacetime containing a traversable wormhole is given by an equation of the form Eq. (\ref{eq:diraccurve2}) where the non-hermitian potential is :
\begin{equation}\label{eq:nonhermitianwh}
V(x)=-i\,\frac{b_0^2}{2 x(b_0^2+x^2)}.
\end{equation}
Notice that in the singularity at $x=0$, the equation is not well-defined, due to the fact that $\Omega=0$. However, we can analyze this equation at both sides of the wormhole's throat. As is well-known in the wormhole literature, this is just an apparent singularity which can be removed by a coordinate change \cite{morristhorne2} and indeed the shape function is chosen in order to allow that physical particles are able to traverse the wormhole \cite{morristhorne,morristhorne2,geodesics,
taylor,qswh}. Moreover, with our choice of coordinates, the spacetime in Eq. (\ref{eq:example}) is just a Minkowski spacetime but with a spatial dependence in the speed of propagation of massless particles. The apparent singularity is just a point at which the velocity of massless particles is 0, which of course does not imply that they stop at this point, since the velocity is finite at both sides of the throat -- as discussed as well in \cite{qswh}.

Finally, we obtain:
\begin{equation}\label{eq:phasewh}
\int{\frac{\Omega'}{\Omega}dx}=\frac{1}{2}\log{\frac{x^2}{b_0^2+x^2}}.
\end{equation}
 
Thus using our results above, we find that any solution of the free 1+1 dimensional massless Dirac equation can be transformed into a solution of the Dirac equation in a spacetime containing a traversable wormhole by means of a suitable local phase transformation, which using Eq. (\ref{eq:densities}) takes the particular form for the probability densities:
\begin{equation}\label{eq:probabilitieswh}
|\psi|^2= \frac{\sqrt{b_0^2+x^2}}{x}|\phi|^2.
\end{equation}

In order to illustrate the power of this technique, we consider wavepacket solutions $\phi(x,t)$ of the free massless 1+1 dimensional Dirac equation in flat spacetime -like the ones that are implemented, for instance in trapped-ion simulators \cite{zitt2}. These can be obtained from an initial wavepacket $\phi(x,0)$ by \cite{thaller}:
\begin{equation}\label{eq:wavepack}
\phi(x,t)=\int_{-\infty}^{\infty} (\phi^+(k)u_{pos}(k,x,t)+\phi^-(k)u_{neg}(k,x,t))dk
\end{equation} 
where $\phi_+(k)$ and $\phi_-(k)$ are the positive and negative parts respectively of the Fourier transform $\phi(k)$ of the initial wavepacket and $u_{pos}(k,x,t)$, $u_{neg}(k,x,t)$ are positive and negative-energy plane-wave solutions respectively. $\phi^+$ and $\phi^-$ are obtained via the inner products $\phi^{\pm}(k)=<u_{pos,neg},\phi(k)>$. In the massless case, the spinors $u_{pos}$ and $u_{neg}$ take a particularly convenient form $u_{pos,neg}=\pm 1/\sqrt{2} (1\, 1)^T$.

In particular, choosing an initial Gaussian wavepacket centered around $x_0$ with width $\sigma$:
\begin{equation}\label{eq:initwavepack}
\phi(x,0)=\frac{1}{\sqrt{N}}e^{-\frac{(x-x_0)^2}{\sigma^2}}\begin{pmatrix}1\\1\end{pmatrix}
\end{equation}
-where $N$ is a normalization constant $N=\sqrt{2\pi\sigma^2}$- we find the following wavepacket solution to the free equation:
\begin{equation}\label{eq:evolvwavepack}
\phi(x,t)=\frac{1}{\sqrt{N}}e^{-\frac{(t-(x-x_0))^2}{\sigma^2}}\begin{pmatrix}1\\1\end{pmatrix}.
\end{equation}

In Fig. \ref{fig:plots} we plot both the free wave packet probability density dynamics in flat spacetime given by the $\phi(x,t)$ in Eq. (\ref{eq:evolvwavepack}) and the corresponding curved spacetime probability density obtained via Eq. (\ref{eq:probabilitieswh}) for different values of the initial position $x_0$. We see that if the wavepacket is initially at the left side of the throat ($x_0<0$) the Dirac particle is able to go through the wormhole, in such a way that the probability density is intensely focused around the throat. On the other hand, if the wavepacket is initially centered around an initial position at the right side of the wormhole throat ($x_0>0$), we see that the probability density gets distorted with an intensity inversely proportional to the initial distance to the throat, as can be seen by comparing Figs. d) and f) with c) and e).

\section*{Discussion}

In this way we are finding for the first time analytical solutions of the dynamics of Dirac particles in a spacetime containing a traversable wormhole. Perhaps more importantly, our technique can be applied in a straightforward fashion to obtain solutions of the Dirac equation in a wormhole background out of numerical solutions of the free Dirac equation in flat spacetime and  out of the experimental data obtained from the various multiplatform quantum simulators of the Dirac equation -without the need of engineering extra terms in the Hamiltonian.

In summary, we have shown how to transform a 1+1 D Dirac equation in curved spacetime into  a flat-spacetime one. The technique consists in applying a suitable local phase transformation to solutions of the flat-spacetime equation. The phase depends on the particular spacetime considered. We show that the dynamics of the transformed Dirac spinor under the full equation - which includes the non-hermitian potential introduced by the curved spacetime metric- can be obtained from the dynamics of the original flat-spacetime solution. In this way, we encode a curved spacetime into a phase. As a first example of what can be done with our technique, we consider a particular case of interest -a family of 1+1 D spacetimes containing a traversable wormhole. In this way, we obtain analytical solutions of the dynamics of a Dirac particle going through a wormhole. Moreover, any available quantum simulator of the Dirac equation would be immediately able to incorporate the analysis of curved spacetimes into the simulation by means of our technique, without the need of implementing extra terms in the Hamiltonian. Thus we expect that our results will be useful for the multidisciplinary community interested in the quantum simulation of relativstic quantum mechanics.

\section*{Acknowledgements}
Financial support by Fundaci\'on General CSIC (Programa ComFuturo) is acknowledged. Additional support by
MINECO Project FIS2015-70856-P (cofinanced by FEDER funds) and CAM PRICYT Project QUITEMAD+
S2013/ICE-2801 is acknowledged.

\section*{Additional information} 
\subsection*{Competing financial interests}
The author declares no competing financial interests.

\end{document}